\documentclass{new_tlp}
\usepackage{graphicx}
\usepackage{balance}  % for  \balance command ON LAST PAGE  (only there!)
\usepackage{mathptmx}
\let\proof\relax

\usepackage{amsthm}
\usepackage{amssymb}
\usepackage[hyphens]{url}
\usepackage[hidelinks]{hyperref}

\newtheorem{example}{Example}

 \newcommand{\comment}[1]{}
 \newcommand{\eat}[1]{}
 \usepackage{listings}
 \usepackage{graphicx}
 \usepackage{xspace}
 \usepackage{soul}
 \def\mt{\tt}
 \def\1S{$\tt _{1S}$}
 \def\+1{{+}1}

 \newcommand{\bldl}{\smallskip\[\begin{array}{ll}}
 \newcommand{\cldl}{\[\begin{array}{ll}}
 \newcommand{\eldl}{\end{array}\]\rm}
 \newcommand{\prule}[2]{ \mt #1 \leftarrow & \mt #2 \\}
 
 \def\pfact#1#2{\mt  #1 &  \mt #2 \\}
 \def\pbody#1#2{ \mt #1 & \mt #2 \\}
 \def\inv{\vspace{-0.2cm}}

 \def\prem{$\cal P$$\!reM$\xspace}

 \def\t3arrow{--\hspace{-3.0ex}^{top3}\hspace{-2.6ex}--\hspace{-2.96ex}\rightharpoondown}

 \def\rof#1{$\tt r_{#1}$}

\begin{document}
\title{Monotonic Properties of Completed Aggregates in Recursive Queries}
\author[~~]
{Carlo Zaniolo,$\;$Ariyam Das,$\;$Jiaqi Gu,$\;$Youfu Li,$\;$ Mingda li, $\;$ Jin Wang\\
{ \normalsize  University of California at Los Angeles}\vspace{-0.1ex}\\
\email{\small  \{zaniolo,ariyam,victor2001,youfuli,limingda,jinwang\}@cs.ucla.edu}\vspace{-2ex}}
% youfuli, jinwang, limingda
%\titlerunning{ Aggregates for Big Data Algorithms }

%%%%%%%%%%%%%%%%%%%%%%%%%%%%%%%%

\maketitle
\begin{abstract}
The use of aggregates in recursion enables  efficient and scalable support
for a wide range of BigData algorithms, including those used in graph applications, KDD applications, and  ML 
 applications, which have proven difficult to be expressed and supported efficiently in BigData systems supporting Datalog or SQL. 
 The problem with these languages and systems is that, to 
avoid the semantic and computational issues created by non-monotonic
constructs in recursion, they only allow programs that are 
stratified with respect to negation and aggregates. Now, while this crippling restriction
is well-justified for negation, it is frequently unjustified 
for aggregates, since  (i) aggregates are often monotonic in the standard lattice of set-containment, (ii)
the \prem property guarantees that
programs with extrema in recursion  are equivalent to stratified programs where extrema are used 
 as post-constraints, and (iii) any
 program computing any aggregates on sets of facts of predictable cardinality tantamounts to stratified programs where the pre-computation 
of the cardinality of the set is followed by a stratum where recursive rules only use monotonic constructs. With (i) and (ii) covered in previous papers, this
paper focuses on (iii) using examples of great practical interest.  For such examples,
we provide a formal semantics that is conducive to efficient and scalable implementations
via well-known techniques such as semi-naive fixpoint currently supported by most Datalog and SQL3 systems.

\end{abstract}

%\clearpage
\section{Introduction}
The growth of BigData applications adds new vigor  to the quest by
Datalog  researchers  for combining the  expressive power  
found  in  recursive Prolog programs  with the  performance and 
scalability of relational DBMSs. 
  Their research led to the delivery of a
   first commercial Datalog system \cite{aref2015design} and 
  impacted significantly  query languages and systems. 
 In particular, many DBMS vendors introduced support for recursive queries into their 
 systems  and the SQL-2003 standards  by 
 adopting  Datalog's key notions and techniques, including  (a) stratified semantics for 
 negation and aggregates, (b) optimization techniques that include (i) semi-naive
  fixpoint computation, (ii) constraint pushing for left/right-linear rules, and (iii)
  magic sets for linear rules.  
However, the impact of these extensions in the market place was  limited,
particularly if we compare it with OLAPs and Data Cubes of descriptive analytics which,
by providing simple extensions of SQL aggregates, allowed DBMS vendors to make
very successful inroads into the brave new world of Big Data analytics \cite{DBLP:journals/sigmod/ChaudhuriD97}.
However, as Big Data analytics grew in diversity and complexity, DBMS showed their
limitations by their inability to support KDD, graph, and ML applications. For instance, (I) Agrawal et al. \cite{DBLP:conf/sigmod/SarawagiTA98} showed that KDD applications are very  difficult to support in SQL DBMS,  while (II) 
 Stonebraker et al. \cite{Stonebraker:2010:MPD:1629175.1629197}  pointed out that MapReduce owed its 
 extraordinary popularity to its success in extension to recursive applications, such as Page Rank, 
 but they never called for extensions of data parallelism of SQL DBMS that could support such new applications. 
 Finally, (III) more than a dozen of graph database systems have been developed for this important application domain
 that is problematic for SQL DBMSes.
%
%  Yet,  it soon became clear that even with these extensions supporting
%  data mining algorithms, such as Apriori, was beyond the capabilities of SQL or other query 
%languages \cite{}. While a new version of Apriori \cite{xxx} made it clear 
%  that this only uses  continuous count aggregate, which being monotonic 
%in the lattice of set containment, could be allowed freely in recursion this
%as far as we know, this useful extension of SQL 
%was not explored in commercial DBMS.     Also after map-reduced was
%introduced to support the parallel computation of page-rank, DB researchers
%were quick to point out correctly that this kind of data-parallelism  had been
%supported by SQL systems for 20 or so year,  but they did no mention the need
%forSQL  extensions that could support massive graph queries such
%as the recursive ones required for page-rank computation.   Yet, with 12
%or so graph DBs being proposed it is only natural that we wonder if there
%are simple extensions of SQL that can be effective on graph applications,
%and similar questions come when it comes to KDD and ML applications.
Now if we examine the many factors that led to this non-optimal situation, we see that
unresolved research issues played a role of paramount importance. 
For instance, since the early days of Datalog, researchers had been aware of the fact that many algorithms can be expressed quite naturally in Datalog once aggregates, and non-monotonic constructs, such as choice, were allowed in recursion, as per the following
 incomplete citation list:  \cite{DBLP:conf/vldb/MumickPR90,DBLP:conf/slp/KempS91,DBLP:conf/pods/GangulyGZ91,DBLP:conf/vldb/SudarshanR91} and \cite{DBLP:conf/pods/GrecoZG92,ross1992monotonic,van1993foundations,DBLP:journals/jcss/GangulyGZ95}.
 But as we will discuss later in the `related work' section, 
 these early proposals suffered from  various limitations,  and
non-monotonic reasoning research was still evolving discouraged premature commitments to a particular solution. 
In the years since then, we have seen major progress on the semantic with the stable-model semantics \cite{Gelfond88thestable} gaining widespread acceptance.
This is due to its great power and generality that extends beyond the original focus of Datalog to cover disjunctive programs and answer-set semantics.
Unfortunately, in its general form, answer-set semantic requires computational complexity levels that are unsuitable for the BigData
applications that have become the computer science cynosure.
As a result, the approaches for non-monotonic semantics that enables efficient implementations for a very wide range of applications has now become an overwhelming need.
This is because, in recent years,
Big-Data applications for graphs, KDD and ML  have grown by leaps and bounds,
laying bare the inadequacy of stratified Datalog and SQL in such crucial domains. 

Many systems have addressed this surge of critical BigData applications by providing specialized libraries
of functions that are written in various PLs which either operate externally 
on data extracted from the DBMS, or internally as an extension of DBMS. 
While, this approach can be effective for some applications and systems, a growing number of researchers have been
investigating whether it is possible to extend to these new applications the old DBMS paradigm, in which usability, portability, and scalability were achieved by writing high-level queries that 
the system then executes efficiently using sophisticated query optimization and data parallelism techniques
 \cite{seo2013distributed,mazuran2013declarative,wang2015asynchronous,bigdatalog,DBLP:journals/vldb/YangSZ17,rasql,appl-iclp19,ssp-iclp19}.
All these research projects share important common points, including (i) the use of parallel, multi-processor or
multi-node, architectures to achieve performance and scalability and
 (ii) the use of aggregates in recursion to express more advanced applications.
However the UCLA projects also address the difficult semantic issues raised by 
aggregates in recursion, which were left untouched by the other projects. 
Indeed UCLA's work shows that formal semantics can be combined with generality and superior performance, and
and even enhance it in unexpected ways. 
In fact, the Pre-Mappability property that was introduced to achieve formal semantics can often deliver better performance by
compensating for skewness across wokers in a parallel execution \cite{ssp-iclp19,DBLP:conf/cikm/DasGZ18}.
%\cite{seo2013distributed,mazuran2013declarative,wang2015asynchronous,bigdatalog,DBLP:journals/vldb/YangSZ17,rasql,appl-iclp19,ssp-iclp19}.

Previous works have shown various ways in which aggregates can be used in recursive logic programs 
while retaining formal semantics. 
Thus, using aggregates that are monotonic in the lattice of set containment  was discussed
in \cite{mazuran2013declarative,bigdatalog,DBLP:journals/vldb/YangSZ17}.
Among the aggregates, using min and max in recursive programs that are equivalent to stratified programs was discussed in \cite{DBLP:journals/tplp/ZanioloYDSCI17,rasql}. 
In this paper, we explore a third important situation where programs using non-monotonic aggregates nevertheless define monotonic mappings because those aggregates are  applied to sets of known cardinalities.
In fact the computation of an  aggregate such as sum 
is performed in two phases.  In the initial phase, 
we progressively add to the 
current  continuous sum  each item in the set. In the final
phase we detect the end of the input and return the last value
produced  in the initial phase. The desirability of clearly
distinguishing between the
two phases is  well- recognized when dealing with
continuous queries on data streams:  in fact, aggregates
returning only the results from initial phase computation are non-blocking, whereas
those returning results produced in the secon phase  are blocking.
Likewise, the need to
provide users with continuous aggregates that only
compute in their initial phase was recognized in SQL:2003
 with the introduction of OLAP Functions
that support continuous aggregates. Drawing a clear distiction
between the continuous and final version of the  aggregate
 produced in the two phases, is also very important when
 using them in recursion. Indeed, the continuous initial aggregates
are monotonic, whereas the final ones are non-monotonic and
they cannot be used as such in recursion.
However, in many situations, including those
where the cardinality of the set is known, the
the final phase computation can be recast in  monotonic terms,
whereby the whole aggregate becomes monotonic
and can be used to express concisely and efficiently powerful recursive queries,
as discussed in the rest of the paper, which is organized as follows.
In the next section we formally define the initial and final versions of
aggregates and the notion of pre-computability for the latter.
In Section 4, we extend these notions to group-by aggregates,
and show how they make possible the simple expression and 
efficient computation Markov Chains, and
Lloyd's clustering algorithm.  Then, the conclusion, in Section 5, 
points out that many KDD and ML algorithms can be expressed
in a similar way.

\section{Defining Set Aggregates}
We begin by defining a general template  to
compute aggregates using Horn Clauses.
More specifically,  let $\tt p(X)$ denote a set of atoms (no duplicates), i.e,
 $ \Gamma= \tt \{ X| p(X) \}$.
Now, the {\em continuous count} aggregate on $\Gamma$ returns the set of positive integers that do not exceeds the cardinality of
the set $ \Gamma$ and can
be computed as follows using Horn clauses:

\begin{example} [Defining continuous count]
\cldl
\prule{r_1: ccp(C, [X])}{p(X), C=1.}
\prule{r_2: ccp(C1, [X|S])}{ p(X),  ccp(C, L),  C1=C+1,   new(X, L).}
\prule{r_3: new(X, [Y|L])}{X <>Y, new(X, L).}
\pfact{r_4:new(X, [~]).}{}
\eldl
\end{example}
Thus  the goal $\tt ? ccp(X)$ progressively returns integers up the actual cardinality of the
above set $\Gamma$.
The name {\em monotonic count} is also used for continuous count, since it is defined using by Horn clauses that always generate monotonic mappings in the lattice of set-containment.

In terms of implementation, the above formal definition of monotonic count is quite inefficient since it constructs  all possible permutations of the $X$ values, while only one of such permutations needs to be considered.
% Given that all others deliver the same result at each step of the computation. 
Thus, actual realizations of  continuous count in systems  visit each atom in $ S$ in  some efficient way---typically in the sequential order in which the atoms are stored.

The traditional {\em final count} used in SQL-2 i.e. the cardinality of set $S$,
can be derived as the maximum of the continuous count $\tt ccp$. 
But rather than using this approach that defines one aggregate using another,
 we can define it by the rules  \rof{1}, \rof{2},  \rof{3} and \rof{4} of Example 1 and the following final rule:
 
 \begin{example} [Defining Final Count from continuous count in Example 1]
\label{ex:fcount}
\cldl
\prule{r_5: final\_count(C)}{ ccp(C,\_), C1=C+1, \neg  ccp(C1, \_).}
\eldl
\end{example}

Thus the final count is defined using negation, and therefore it is non-monotonic,  unlike continuous count. 
But as in the case of continuous count, its implementation will be expedited by considering only one of the possible permutations
of the values in $S$.  
Furthermore, the rule \rof{5} condition $\gamma:\tt ccp(C, \_), C1=C+1,  \neg ccp(C1, \_)~$ can be implemented  by any test that determines that this is the last atom in the set.
For instance, if  the $\tt p(X)$ facts are stored in a file, then the logical condition $\gamma$ is realized by detecting that the next datum is the end-of-file  ($\tt EoF$) mark. 
This is just one way in which the realization of condition $\gamma$ is detected in DBMS.  
Indeed, if our table is indexed using a B+ tree, then there might not be any $\tt EoF$ mark in the data blocks,
and the termination condition is realized by the last bottom level index block with a null pointer to the next block. 
Moreover, if $\tt p(X)$ is the result of a join or other relational algebra expressions, it is the responsibility of the DBMS to signal to the function implementing the aggregate that the $\gamma$ condition is satisfied because all the data satisfying the expression has been generated.
%By providing special constructs for aggregates such as count or sum,  Datalog and SQL query languages
%enhance the ease of use while paving the way to the critical optimization just discussed.
To determine the semantic properties of aggregates, we still need to
consider both (i)  their explicit
logic-based definitions, such as the one just described for count,
and (ii) the fact that the  completion condition $\gamma$ might not be part of the logic program
expressing the application at hand since it is implicitly tested by the underlying system.
%In actual implementations the computation of count is expedited  only one
%permutation needs to be consider as they all produce the same +1 result.
%Also in the storage structure in actual implementations provide a special
%EOF character or other means to detect when all the items in the set have
%been used: as such the last value of $\tt C$ before that represent the 
%correct actual count, with no  explicit execution of  rule \rof{3}. While we will
%certainly use these operational improvement in our systems,  in terms of 
%formal semantics we must use the stratified program above based on all-permutation
%of the members in the set a-la slow sort \cite{xxxx}.

\inv\subsection{From Counts to Sums and Averages}
The definition of other aggregates such as sum or count will use the template established for
count  consisting of an initial phase where  their continuous version is computed, and then of a second phase where the final result is returned. 
Moreover the  final count can be used as the completion test that brings 
about the final  phase in the computation of these aggregates.

For instance, the sum of the $\tt X$-values that satisfy  $\tt p(X) \in S$
can be defined as shown in Example \ref{ex:sum} where rules  \rof{6} and \rof{7} compute both the continuous sum and the continuous count
as the first and the second arguments of $\tt csc$.  
The value of the final sum is actually the value of the continuous sum when 
the continuous count value reaches a value that is equal to the cardinality 
of our set $\Gamma$, i.e. a value that is equal to the final count.  Rule 
\rof{8} expresses this completion condition using the predicate $\tt new$ that was defined  in Example \ref{ex:fcount} for the computation of $\tt final\_count$.

\begin{example}[Defining continuous and  final sum]
\label{ex:sum}
\cldl
\prule{r_6: csc(S, C, [X])}{p(X), S=X, C=1.}
 \prule{r_7: csc(S1, C1, [X|S])}{ p(X),  csc(S, C, L),  S1=S+X, C1=C+1, \neg new(X, L).}
 \prule{r_8: final\_sum(S)}{csc(S, C,\_),  C1=C+1, \neg  csc(C1, \_\_).}
 \eldl
 \end{example}
 
Thus, the sum aggregates is basically defined by a monotonic computation, except for a final rule that call on final-count predicate which is non-monotonic.
However, in many situations final-count is known before we enter the recursive computation of $\tt csc$. 
For instance, this is true when $ \Gamma$ represents the atoms of a vector of known length. 
Moreover, in many situation where the cardinality of $\Gamma$ is not known, it can be actually  
computed using the program in Example 1, to produce $\tt final\_count(C)$ 
in a lower stratum.
Then, the rules  \rof{6} and \rof{7} in Example \ref{ex:sum} will still be used to compute $\tt csc$m,  but instead of \rof{8} we will use the following rule to compute the final-sum:
 \cldl
  \prule{r_9: final\_sum(S)}{  csc(S, C,\_),  final\_count(C).}
 \eldl
 
 Therefore, the final sum aggregate can be implemented by a stratified program where the lower stratum perform the non-monotonic computation of final count, and the next stratum derives sum by a monotonic computation since rules \rof{6}, \rof{7} and \rof{9} do not use negation.

 Thus, in our definition of sum we have combined the computation of sum and count
 into one stratified programs, where rules \rof{1}, and \rof{2} occupy a lower stratum,
 the rule \rof{3} containing negation is at higher stratum, and rules \rof{4}, \rof{5}
 and \rof{6} a still higher stratum.   
 Only  rule \rof{3} that determines the actual count is non-monotonic. 
 
 In every program, recursive or not, when we will have to compute sum on a set 
 whose cardinality is already know, rule \rof{3} is no longer needed and
  the computation of our aggregate becomes yet another monotonic predicated defined using Horn clauses.  
  In practice, the situation is even more dramatic since the detection of  the $\tt  EoF$ or other termination condition can be used
 to replace rule \rof{6} {\em provided that the actual count computation could
 have been completed and saved away before the computation of sum}---
 whereby $\tt  EoF$or other completion condition will simply trigger the retrieval of this value to execute rule \rof{6}. 

Similar observations also hold true for other aggregates such as average, and extrema aggregates.
In fact, the average aggregates can be computed by replacing rule  $\tt r_8$ with the following rule:
\cldl
 \prule{ final\_avg(Avg)}{  csp(S, C,\_), Avg=S/C, final\_count(C).} \eldl

For Max, we can instead write the following rules where we use the predicate $\tt larger$ to return the larger of the two values
$\tt M$ and  $\tt X$ (they cannot be equal since we are using set semantics).

\begin{example}[Defining the max on a set where final\_count is known.]
\label{ex:max}
\cldl
\prule{ccs(S, C, [X])}{ p(X), M=X, C=1.}
\prule{cmp(S1, M1, [X|S])}{ p(X),  cmp(M, C, L), larger(M, X, M1), notin(X, L).}
\prule{larger(X, Y, X)}{X>Y.}
\prule{larger(X, Y, Y)}{X\leq Y.}
\prule{ final\_max(M)}{cmp(M, C,\_), final_count(C).}
 \eldl
 \end{example}
Dual definition holds for min, where instead of larger we will use a predicate that returns the smaller of the two values.

\section{Group-By Aggregates}

The logical definition of aggregates specified with a group-by clause 
can be derived as an extension of that provided in the previous section.
Take for instance the following rule:
\cldl
\prule {r_a: qs(X, sum\langle Y \rangle)}{pairs(X, Y).}
\eldl

Then the join computation of sum and count can be performed as follows:

\begin{example} [Defining continuous sum and  final sum in the presence of group-by]
\label{ex:csumgroup}
\cldl
\prule{r_b: gbsc(X,S, C,[~])}{q(X,Y), S=0, C=0.}
 \prule{r_c: gbsc(X, S1, C1, [[X,Y | L])}{ q(X,Y),  gbsc(S, C, L),  S1=S+X, C1=C+1,}
 \pbody{}{ \neg new([X,Y]),  L).}
  \prule{r_d: gbsc(X, S, C1, [[X,Y | L])}{ q(X,Y),  gbsc(S, C, L), ,C1=C+1,}
 \pbody{}{ \neg new([X,Y]),  L).}
 \prule{r_e: final\_sum(S)}{gbsc(S, C,\_),  C1=C+1, \neg  gbsc(C1, \_,\_).}
 \eldl
 \end{example}

Thus, the computation starts with \rof{a} that sets the values of sum and count to zero.
Then, after checking that the pair $\tt [X,Y]$ is in fact new, we increase the 
values of $\tt S$ and $\tt C$ for the group-by value matching $\tt X$, but we only 
increase the  $\tt C$ value for the rest.
Thus at the end of the fixpoint computation,
for each $\tt X$ value we will have the sum $\tt S$  of the $\tt Y$-values associated
with it.  The  $\tt C$ values will be the same for every $\tt X$ since it is
equal to the cardinality of the set containing the $\tt p(X,Y)$ facts.

The final rule \rof{e} returns the  final value of  continuous
sum when the continuous count has reached its  final value.
This is the only rule using negation, and
if we can replace it by conditions expressed using negation the 
whole computation of final  sum becomes monotonic.
This is, for instance, the situation when the value of $\tt C$
can be pre-computed  before we enter into the computation of sum
as in the case in which is computed on a set of fixed or
Pre-Countable Cardinality (PCC),
This logic-based definition  of  group-by sum on PCC sets cardinality is easily other  aggregates. In fact, if in Example \ref{ex:csumgroup} we replace sum by count, avg, min and max,  we obtain a well-defined and efficiently
computable semantics. 
%However, if $\tt pairs$  and $\tt qs$ in \rof{a} were the same predicate or they are mutually recursive, a formal semantics exists only 
%in the important special cases we will discuss next.

\section{Pre-Countable Cardinalities in Recursion}

Many examples of great practical interest belong to the PCC category. 
For instance, the Markov-Chain application is of great interest because it is closely related 
to the Page Rank algorithm that is at root of the Map-Reduce developments.
 % and because  it provides an interesting situation where the sum 
%is applied to a set of known cardinality,
 We assume that we are given as DB facts  
 $\tt mov(City, DestCity,  Perc)$  which respectively
 describe the names of the cities of interest and  the fraction of the  
population that will (most likely) move from $\tt Cit$  to $\tt DesCity$ in the course of a year.  
For each city there is also a non-zero arc from the city back to the same city showing people that will not move away. 
Therefore, the sum of $\tt Perc$  for the  arcs leaving the city (i.e., a node) is  always equal to one. 
 
 Thus, assuming that initially every city has a population of, say, $100 k$, we need to find how the  population evolves over the years.   
 For that we can use the following program:
 
 \begin{example} [The kernel of the Markov Chain algorithm]
\label{ex:markov}
 \cldl
\prule{\!next(0, Cit, sum\langle  In \rangle)} {mov(Cit, Cit, \_ ), In= 100000.}
\prule{\!next(J1, To, sum\langle  In \rangle)}{next(J, Cit, Pop), mov(Cit, To, Perc),}
\pbody{} {In = Pop {\times} Perc, J1{=} J{+}1).}
\eldl
\end{example}
Now observe that, in the course of the fixpoint computation, the atoms with index value $\tt J1=J{+}1$  are generated after those with index value  $\tt J$. 
Thus, the computation of  $\tt next(J1, Cit, sum\langle  In \rangle)$ will return
the current value of the continuous sum as its final-sum as soon 
as the computation of $\tt next(J1, Cit,  In)$ completes reaching the final count. 
This final count in fact is equal to the size of results obtained by joining $\tt next(J, Cit, Pop)$ with $\tt mov(Cit, To, Perc)$, which is actually independent from $\tt J$. 
Therefore, this count value can be computed at a stratum lower than the stratum of $\tt next$ by rules defining a predicate named, say, $\tt sharedcount(C)$, which will then be passed to the rules of our program via an additional $\tt sharedcount(C)$ goal added to the rules defining $\tt next$.
The stratified program so obtained defines the formal semantics of our logic program. Now, this formal semantics can be realized by an operational semantics that dispenses from the computation of  $\tt sharedcount(C)$, and simply detects
the completion of the natural join  completion of the natural join  of  $\tt next(J, Cit, Pop)$ with $\tt mov(Cit, To, Perc)$. In fact we count the
number of tuples being produced by the natural join, this will return $\tt C$
upon its completion, whereby the test specified by  \rof{e} can be
replaced by a test that the computation of the join is completed---a test that
is already  built-in the implementation provided by the system.

\paragraph*{Termination and Optimization}
In addition to issues of formal semantics, the approach discussed in the previous section also allows us to deal effectively with   termination and optimization issues.  
For instance, in the current form, our Markov chain example, is non-terminating.
A simple solution to that is to introduce a condition such as $\tt J \leq 10000$ if we want to ensure that the fixpoint iteration terminates in $1000$  steps. 
Typically however, $1000$ steps are not needed for the computation to converge to a state  in which each successive step returns the same $\tt Cit$ and $\tt Pop$ results as the previous step. 
To stop as soon as convergence occurs, an additional goal can be added to check that the population has increased in some city (and therefore decreased in others).
Finally, we might want to specify that we are only interested in the final, i.e., the max value for the index $\tt J$. 
We then obtain the following program:

 \begin{example} [The actual Markov chain algorithm]
 \cldl
\prule{\!next(0, Cit, sum\langle  In \rangle)} {mov(Cit, Cit, \_ ), In= 100000.}
\prule{\!next(J1, To, sum\langle  In \rangle)}{next(J, Cit, Pop), mov(Cit, To, Perc),}
\pbody{} {In = Pop {\times} Perc, J1{=} J{+}1),  J1 \leq 1000,}
\pbody{} {JL=J-1,   next(JL, Cit, PopL), PopL - Pop>0.} 
\prule {finalstep(max\langle J\rangle)}{next(J, \_, \_).}
\prule {fpop(City, Pop)} {finalstep(J), next(J, Cit, Pop).}
\eldl
\end{example}
The last two rules specify post-conditions that must applied at the end of the fixpoint computation; 
however, it is quite straightforward for the compiler to integrate them into the semi-naive fixpoint computation to achieve a significant optimization. 
In fact, during the semi-naive fixpoint computation,  the compiler identifies new facts (i.e., the delta)
obtained at each point in the computation. 
Moreover, the latest delta atoms are identified by the latest value off $\tt J$, 
which is also the max value of the index $\tt ~J,$ i.e., the values  returned by \rof{e} upon termination.
Thus \rof{e} and \rof{f} can be implemented by simply returning the latest delta atoms upon termination~\footnote{An alternative approach using chain max is also available.}.

Now, the fact that only the atoms for the max value of $\tt J$ are needed implies that all others can be dropped to achieve a much more efficient usage of memory.

Therefore, we have now a formal semantics defined by a stratified program
consisting of (i) a bottom stratum where count is defined, (ii) a  middle stratum of   Horn clauses, i.e., monotonic rules , and (iii)
a top stratum used to post-select the final results of interest. Now, this
formal is realized via a very efficient  operational 
semantics that only requires the semi-naive computation in the middle stratum, inasmuch as
the completion of the join in (i) replaces the completion of the final count, and 
the extraction of the final results in (iii) is realized by the selection of the final delta
in the semi-naive fixpoint.

Similar conclusions and optimizations hold for the popular clustering technique
known as Lloyd's algorithm discussed next.

 \paragraph*{ Lloyd's Clustering Algorithm}
 
We are given a large set of K-dimensional points.  
Each point is described by a unique $\tt Pno$ and the  coordinate values in each of its $K$
dimensions, i.e., by   $\tt point(Pno, Dim, Value)$. 
We also have a small set of centroids, e.g., say that we have 10 such points.
Then to generate the initial assignment $\tt center(0, Cno, Dim, Val)$, 
we used the predicate $\tt init(Cno, Dim, Val)$ that implements one of the many techniques described in the literature. 
Therefore, we have algorithm shown in Example \ref{ex:lloyd}. At each step $\tt J$, the
algorithm finds the closest center for each point.
Then a new set can be generated by averaging their coordinates~\footnote{In our rules we use the $\tt encd$ and $\tt decd$ to
represent to short integers by one long integer and to reconstruct the original number}.

 \begin{example} [Clustering a l\'a Lloyd]
 \label{ex:lloyd}
  \cldl
  \prule{r_: center(0, Cno, Dim, Val)}{ init(Cno, Dim, Val).~~~~~~~~~~~~~~~~~~~}
 \eldl
 \inv\inv \cldl
  \prule{r_b: dist(J, Pno,~Cno,~ sum\langle SqDis\rangle)}{point(Pno, Dim, Val) ,  center(J, Cno, Dim, CVal),}
  \pbody{} {SqDis= (Val-Cval)*(Val-Cval).}
\prule{r_c: mindist ( J, Pno,~min \langle DCno \rangle)}{ dist(J, Pno, Cno,  DSm),
 encd(DSm, Cno, DCno).}
\prule{r_d: center(J1, Cno,  Dim, avg \langle Val\rangle )}{mindist(J, Pno, DmCno),
 decd(DmCno, \_, Cno),}
\pbody{}{points( Pno, Dim, Val),  J1=J+1.}
\eldl
\end{example}
Thus, if we let $|P|$, $|C|$ and $|D|$ denote the respective cardinalities of our set of points,
centers, and dimensions.
We have that, for each rule, the aggregate computation involves a number of elements that is independent of $\tt J$: \rof{b} specified a computation taking place over $|P| \times |C| \times |D|$, \rof{c} specified a computation is over 
$|P| \times |C|$ elements, and the computation of  \rof{2}  is over $|P| \times |D|$ elements.
These are counts that can be easily determined before the recursive computation and remain the same for every value of $\tt J$. 
These explicit values could be passed to the recursive rules for computing the monotonic versions of the aggregates used 
in these rules. 
But a much simpler and efficient solution consists in letting the system detect the completion of execution of the body operators at  each step $\tt J$, which
is already implemented  as part of the  optimized seminaive fixpoint computation,
as in the case of the Markov Chain  computation.  Indeed,
issues similar to the termination and optimization that we discussed for Markov Chain also hold for Lloyd's algorithm.

%Too bad that the last two lines could not be combined. 
%Now  we found  the new set of
%center, whereby the algorithm repeats this  basic new computation step. 
%Observe that we do naive.  
% Could we do seminaive instead--some very interesting questions.
% They are the same when we use $J$.
%
%This also apply to classifiers, where tree are systematically extended one level
%after the other, by basically making an additional pass through the dataset.
%
%Observe the difference with a transitive closure where we do not know what 
%new shortest paths might be introduced during the computation. More.

%\input{mutual}
%\input{extended} 
%\input{programming} 
%\input{newrelated} 
\section{Conclusion}
Recursive computations on datasets of fixed cardinality represent an 
area of great theoretical and practical interest for Datalog and other logic-based languages. 
Indeed, we have shown that important applications such Markoff Chains and Lloyd's clustering algorithm can be expressed 
very efficiently using aggregates in recursion, 
while avoiding the difficult semantic issues besetting the use of non-monotonic constructs in recursive programs.
In all these examples, we dealt with facts describing a given set of entities, such as cities, by their attributes (such as population). We have found that, when the number of such entities in the world remains unchanged, aggregates on the attributes of
these entities can can be used in recursive logic rules while preserving the desirable properties of fixpoint computations.  
This and other recent results using the Pre-Mappability property of extrema~\cite{rasql,appl-iclp19,ssp-iclp19} suggest that aggregate
can provide the long-sought bridge between formal non-monotonic semantics and efficient scalable big data computations that, over many years of work, could not be build by non-monotonic reasoning researchers using only negation.
\balance
\clearpage
\bibliographystyle{abbrv}
\bibliography{monotonic,tlp}
\end{document}